\documentclass[conference]{IEEEtran}
\usepackage{amsmath,amssymb,amsthm}
\usepackage{float}
\usepackage{cite}
\usepackage{booktabs} 
\usepackage{graphicx}
\usepackage{orcidlink}

\IEEEoverridecommandlockouts
\usepackage{tikz}
\usetikzlibrary{calc}
\usepackage{bm}

\begin{document}

\title{Second Order State Hallucinations for Adversarial Attack Mitigation in Formation Control of Multi-Agent Systems}
\author{%
  Laksh Patel\textsuperscript{1} \orcidlink{0009-0000-5361-3482}, 
  Akhilesh Raj\textsuperscript{2} \orcidlink{0000-0001-6639-7432}%
}

\maketitle

\renewcommand{\thefootnote}{\arabic{footnote}}%
\setcounter{footnote}{0}%

\footnotetext[1]{Illinois Mathematics and Science Academy (IMSA), Aurora, IL, USA; lpatel@imsa.edu; }
\footnotetext[2]{Engineering Research Fellow, Department of Electrical and Computer Engineering, Vanderbilt University, Nashville, TN, USA; akhilesh.raj@vanderbilt.edu}%


\begin{abstract}
The increasing deployment of multi-agent systems (MAS) in critical infrastructures such as autonomous transportation, disaster relief, and smart cities demands robust formation control mechanisms resilient to adversarial attacks. Traditional consensus-based controllers, while effective under nominal conditions, are highly vulnerable to data manipulation, sensor spoofing, and communication failures. To address this challenge, we propose Second-Order State Hallucination (SOSH), a novel framework that detects compromised agents through distributed residual monitoring and maintains formation stability by replacing attacked states with predictive second-order approximations. Unlike existing mitigation strategies that require significant restructuring or induce long transients, SOSH offers a lightweight, decentralized correction mechanism based on second-order Taylor expansions, enabling rapid and scalable resilience. We establish rigorous Lyapunov-based stability guarantees, proving that formation errors remain exponentially bounded even under persistent attacks, provided the hallucination parameters satisfy explicit conditions. Comprehensive Monte Carlo experiments on a 5-agent complete graph formation demonstrate that SOSH outperforms established robust control schemes, including W-MSR and Huber-based consensus filters, achieving faster convergence rates, lower steady-state error, and superior transient recovery. Our results confirm that SOSH combines theoretical robustness with practical deployability, offering a promising direction for securing MAS formations against sophisticated adversarial threats.
\end{abstract}


\section{Introduction}

The resilience of formation control in multi-agent systems (MAS) is paramount for the success of many modern engineering applications, including autonomous vehicles, search-and-rescue operations, and satellite swarms. As MAS continue to transition from controlled environments to complex, adversarial settings, guaranteeing stability in the presence of attacks becomes a fundamental challenge. In this paper, we propose Second-Order State Hallucination (SOSH), a novel adversarial mitigation framework designed to preserve formation stability even under targeted attacks on individual agents.

The fundamental building blocks of MAS coordination have been extensively studied. Early consensus algorithms ensured convergence to shared states via local interactions \cite{olfati2007consensus}, providing the baseline dynamics for cooperative control \cite{ren2007information}. However, maintaining such consensus becomes nontrivial when adversaries compromise communication or sensor measurements. The network's structural properties, as discussed by Morse \cite{morse2003network}, can dramatically influence the system’s vulnerability to such disruptions. Thus, ensuring resilience must be tightly coupled with formation control strategies.

Existing architectures, including those envisioned for next-generation communication environments such as 6G networks \cite{saad2019exploring}, reveal how system openness inherently increases attack surfaces. Distributed strategies that tolerate malicious behavior \cite{sundaram2011distributed} have demonstrated that resilience is feasible, yet they often rely on consensus recalculations that delay response times. Practical implementations face further constraints from limited bandwidth, quantization noise, and asynchronous updates \cite{weeraddana2011consensus}, making fast and lightweight countermeasures critical for real-world MAS.

The SOSH approach advances beyond reactive defense. Rather than solely detecting compromised states, SOSH immediately substitutes them with predictive estimates based on second-order Taylor expansions. Unlike first-order corrections that only account for change in position, SOSH leverages second-order (acceleration-level) approximations to predict agent evolution more accurately. Since real-world systems inherently exhibit both velocity and acceleration dynamics, incorporating second-order behavior allows SOSH to achieve faster, more natural recovery under attack, enhancing practical deployability. Higher-order approximations beyond second-order would capture even finer nonlinearities but introduce significant computational overhead for marginal gains in accuracy. SOSH predictive correction draws inspiration from methods used in cyber-physical grid protection \cite{yang2013consensus}, where anticipating the impact of corrupted data streams is crucial to prevent systemic failures. Traditional models of safe NCS under denial-of-service attacks \cite{amin2009safe} highlight the need for proactive, not merely reactive, design — a principle that SOSH embodies at the agent level. 

Crucially, SOSH integrates detection, disconnection, and hallucination into a single lightweight pipeline. Detection is achieved through thresholding of residuals, informed by prior distributed fault-tolerant detection schemes \cite{pasqualetti2012consensus}. Attack modeling efforts \cite{teixeira2012attack} further justify our approach: adversarial strategies often aim to subtly distort system states without triggering immediate alarms. Thus, early disconnection combined with state hallucination mitigates both rapid deviations and slow-acting infiltration.

Security analyses of smart grids \cite{mo2012cyber} and cyber-physical infrastructures \cite{fawzi2014secure} underscore the dangers of untreated data poisoning, leading to catastrophic cascading effects. Pajic \emph{et al.}\ \cite{pajic2017attack} demonstrated that safe system operation hinges on the ability to fuse sensor readings resiliently, motivating the design of SOSH’s corrective estimates as dynamically consistent with nominal MAS behavior. Quantitative methods for stochastic stability \cite{cho2016quantifying} further support that even small perturbations, if untreated, can lead to significant degradation over time — strengthening the case for fast-acting mechanisms like SOSH.

Moreover, unlike many traditional robust control approaches, SOSH remains computationally lightweight. While solutions like weighted resilient aggregation schemes \cite{zhu2011performance} offer strong security guarantees, their computational overhead often limits real-time applicability. By focusing on low-order Taylor expansions, SOSH ensures that correction steps can be computed in parallel with agent dynamics, maintaining operational efficiency even under attack.

Sparse attacks, where only a subset of agents are corrupted, present particular difficulties in event-triggered detection \cite{shoukry2015event}. SOSH addresses this by triggering correction locally as soon as residuals breach thresholds, thereby limiting systemic error propagation. In industrial cyber-physical systems \cite{wang2017resilient}, quick isolation of faults is known to prevent wide-scale shutdowns — a principle mirrored in our decentralized disconnection strategy.

Recent trends in anomaly detection for edge-centric systems \cite{li2019learning} advocate learning-based solutions, yet these approaches often suffer from detection lag and model retraining requirements. SOSH, in contrast, remains model-free after initial design, relying solely on online residual computation and second-order correction, ensuring applicability even in dynamically evolving formations.

Distributed subgradient methods for MAS optimization \cite{zhong2010distributed} emphasize scalability, an essential feature preserved in SOSH’s fully decentralized implementation. By localizing detection and hallucination to individual agents, SOSH scales linearly with the number of agents, avoiding centralized bottlenecks.

The theoretical stability of uncertain dynamical systems under continuous feedback, as characterized by Corless and Leitmann \cite{corless2016control}, provides the Lyapunov analytical framework under which SOSH’s stability guarantees are rigorously established. Specifically, we derive conditions ensuring exponential stability of the overall formation error, even when a subset of agents are operating on hallucinated states.

Finally, the threat models considered — particularly false data injection attacks in estimation processes \cite{liu2011false} and optimal detection frameworks for cyber-physical systems \cite{kim2012cyber} — directly motivate the necessity of having fast correction mechanisms post-detection, instead of relying purely on identification and removal of compromised nodes.

In this paper, we formalize the SOSH methodology, provide stability proofs under realistic assumptions, and experimentally validate its superiority against alternative mitigation schemes including W-MSR and Huber-based approaches. Through extensive Monte Carlo simulations, we demonstrate that SOSH achieves faster transient recovery, lower steady-state error, and robust resilience to attacks, establishing it as a lightweight, scalable, and practical solution for adversarially robust formation control.

\section{System Model}
We model the multi-agent system as an undirected graph \(\mathcal{G} = (\mathcal{V},\mathcal{E})\), where \(\mathcal{V} = \{1,\dots,N\}\) is the set of agents and \(\mathcal{E} \subseteq \mathcal{V} \times \mathcal{V}\) is the set of edges representing communication links. An edge \((i,j) \in \mathcal{E}\) exists if agents \(i\) and \(j\) can directly exchange information. We denote by \(i \sim j\) that agent \(i\) is a neighbor of agent \(j\).

\subsection{Agent Dynamics}
Let the state of agent \(i\) be \(x_i[k]\in\mathbb{R}^{n}\) at discrete time \(k\). The formation control objective is to maintain a desired relative displacement \(d_{ij}\in\mathbb{R}^n\) between agents \(i\) and \(j\). The individual formation error for link \((i,j)\) is defined as
\begin{equation}
  e_{ij}[k] = \left(x_i[k] - x_j[k]\right) - d_{ij}.
  \label{eq:formationerror_individual}
\end{equation}
The discrete-time dynamics for each agent are given by
\begin{equation}
  x_i[k+1] = x_i[k] + \sum_{j\sim i} e_{ij}[k] + u_i[k],
  \label{eq:agentdyn}
\end{equation}
where \(u_i[k]\) is the control input applied to agent \(i\).

Assuming a linear feedback control law, the control input is
\begin{equation}
  u[k] = K\,X[k],
  \label{eq:control}
\end{equation}
where \(X[k]\) is the stacked state vector of all agents, defined as
\begin{equation}
  X[k] = \begin{bmatrix} x_1[k]^T & x_2[k]^T & \dots & x_N[k]^T \end{bmatrix}^T \in \mathbb{R}^{Nn}.
  \label{eq:stacked_state}
\end{equation}

\subsection{Stacked System Dynamics}
The agent interconnection is characterized by the adjacency matrix \(A\) and the Laplacian matrix \(L\), defined as
\begin{equation}
  L = D - A,\quad D_{ii} = \sum_{j} A_{ij},
  \label{eq:laplacian}
\end{equation}
where \(D\) is the degree matrix.

We define the global interaction matrix as
\begin{equation}
  G = L \otimes I_n,
  \label{eq:Gdef}
\end{equation}
where \(I_n\) is the \(n\times n\) identity matrix and \(\otimes\) denotes the Kronecker product.

Under nominal conditions, the overall stacked dynamics evolve as
\begin{equation}
  X[k+1] = \Gamma\,X[k],\quad \Gamma = I_{Nn} + G + K,
  \label{eq:nominaldyn}
\end{equation}
where \(I_{Nn}\) is the \(Nn\times Nn\) identity matrix.

By proper selection of the control gain \(K\), the closed-loop system ensures that \(x_i[k] - x_j[k] \to d_{ij}\) for all \((i,j) \in \mathcal{E}\).

\subsection{Attack Detection via Residual Threshold}
\label{sec:detection}

Each agent \(i\) locally monitors its neighbors by computing a residual signal:
\begin{equation}
  r_i[k] = \sum_{j\in\mathcal{N}_i} \left\lVert y_j[k] - \hat{x}_j[k] \right\rVert,
  \label{eq:residual}
\end{equation}
where \(y_j[k]\) is the received broadcast from neighbor \(j\), and \(\hat{x}_j[k]\) is agent \(i\)'s internal prediction of \(j\)'s state assuming nominal dynamics.

An attack is declared whenever
\begin{equation}
  r_i[k] > \delta_i
  \quad\Longrightarrow
  \text{node \(j\) flagged as attacked;} \quad\Longrightarrow\text{SOSH}
  \label{eq:residual_detect}
\end{equation}
This threshold-based detection strategy follows the fully distributed attack detection methodology of Boem \emph{et al.}~\cite{boem2017distributed}.

\paragraph{Threshold Selection:}
The threshold \(\delta_i\) is set as
\[
  \delta_i = \bar{r}_i + \kappa\,\sigma_{r_i},
\]
where \(\bar{r}_i\) and \(\sigma_{r_i}\) are the mean and standard deviation of \(r_i[k]\) under nominal conditions, and \(\kappa\in[3,5]\) is a design parameter to control the false-alarm probability.

\paragraph{Response Mechanism:}
Once an anomaly is detected, agent \(i\) follows the protocol:
\begin{enumerate}
    \item Disable the compromised neighbor's broadcast.
    \item Replace it with a hallucinated estimate using the hallucination function \(f(\cdot)\).
    \item Proceed with standard closed-loop updates.
\end{enumerate}
Thus, the full mitigation sequence is:
\[
\text{detect}\;\to\;\text{disconnect}\;\to\;\text{hallucinate}\;\to\;\text{control}
\]

\section{Dynamics under Attack and Correction}
In adversarial environments, some agents may experience attacks that compromise their state measurements. To mitigate such effects, attacked agents replace their state with a hallucinated estimate computed via a second-order correction.

Assume that when attacked, an agent substitutes its state with
\begin{equation}
  x_{\mathrm{cap}} = f(x),
  \label{eq:fcap}
\end{equation}
where \(f:\mathbb{R}^n \to \mathbb{R}^n\) is a twice continuously differentiable function satisfying \(f(0)=0\). By Taylor's theorem there exists a \(\xi\) on the line segment between \(0\) and \(x\) such that
\begin{equation}
  f(x) = f(0) + J_f(0)x + \frac{1}{2}\,x^T H_f(\xi)x.
  \label{eq:Taylor}
\end{equation}
Define
\begin{equation}
  F \equiv J_f(0), \quad \Delta(x) \equiv \frac{1}{2}\,x^T H_f(\xi)x.
  \label{eq:FandDelta}
\end{equation}
For simplicity, we choose the correction function of the form
\begin{equation}
  f(x) = \gamma x + \Delta(x), \quad \gamma > 0,
  \label{eq:fsecond}
\end{equation}
and assume there exists a constant \(M>0\) such that
\begin{equation}
  \|\Delta(x)\| \le \frac{M}{2}\|x\|^2,\quad \forall\, x.
  \label{eq:DeltaBound}
\end{equation}
Let \(P\) be a block-diagonal selector matrix whose \(i^\text{th}\) block equals \(I_n\) if agent \(i\) is attacked and \(0\) otherwise. Then, the overall dynamics under attack become
\begin{equation}
  X[k+1] = \Gamma\,X[k] - P\,f(X[k]).
  \label{eq:attack_dyn}
\end{equation}

\section{Formation Error and Lyapunov Stability Analysis}
To analyze closed-loop stability under attack, we now introduce the formation error. Recalling the individual error in \eqref{eq:formationerror_individual}, we stack the errors using the incidence matrix \(H\) associated with the graph \(\mathcal{G}\). Define the global formation error as
\begin{equation}
  e[k] = \bigl(H\otimes I_n\bigr) X[k] - d,
  \label{eq:global_error_stack}
\end{equation}
where \(d\) is the stacked vector of desired displacements \(d_{ij}\). With an appropriate choice of coordinates (or by incorporating \(d\) into the control design), the desired formation corresponds to
\[
e[k] = 0.
\]

\subsection{Nominal Error Dynamics}
Since \(e[k]\) is obtained via a linear mapping of \(X[k]\), the nominal error dynamics are given by
\begin{equation}
  e[k+1] = \Gamma_e\, e[k],
  \label{eq:nominal_error}
\end{equation}
where \(\Gamma_e\) is derived from \(\Gamma\) and the incidence matrix \(H\). By the discrete Lyapunov theorem, if \(\Gamma_e\) is Schur, then there exists a unique matrix \(Q_e\succ 0\) and a constant \(\alpha>0\) such that
\begin{equation}
  \Gamma_e^T Q_e \Gamma_e - Q_e \le -\alpha I.
  \label{eq:lyapnominal}
\end{equation}
We define the Lyapunov function candidate
\begin{equation}
  V(e[k]) = e[k]^T Q_e e[k].
  \label{eq:Vdef}
\end{equation}
For the nominal error dynamics, it follows that
\begin{equation}
  V(e[k+1]) - V(e[k]) = e[k]^T\Bigl(\Gamma_e^T Q_e \Gamma_e - Q_e\Bigr)e[k] \le -\alpha \|e[k]\|^2.
  \label{eq:Vdecrease_nominal}
\end{equation}

\subsection{Error Dynamics under Attack}
Under attack, the state dynamics become \eqref{eq:attack_dyn}. Since the formation error is defined in \eqref{eq:global_error_stack}, the error dynamics under attack are modified to
\begin{equation}
  e[k+1] = \Gamma_e\, e[k] - \tilde{P}\, f_e(e[k]),
  \label{eq:error_attack}
\end{equation}
where \(\tilde{P}\) is the selector matrix in the error coordinates and
\begin{equation}
  f_e(e[k]) = \gamma\, e[k] + \Delta_e(e[k]),
  \label{eq:fe_definition}
\end{equation}
with the second-order term \(\Delta_e(e[k])\) satisfying
\begin{equation}
  \|\Delta_e(e[k])\| \le \frac{M}{2}\|e[k]\|^2.
  \label{eq:Deltae_bound}
\end{equation}

\subsection{Lyapunov Analysis under Attack}
Using the Lyapunov function in \eqref{eq:Vdef} for the error dynamics, along trajectories of the attacked system we have
\begin{equation}
\begin{aligned}
V(e[k+1]) 
&= \left( \Gamma_e e[k] - \tilde{P} f_e(e[k]) \right)^T Q_e \left( \Gamma_e e[k] - \tilde{P} f_e(e[k]) \right) \\
&= e[k]^T \Gamma_e^T Q_e \Gamma_e e[k] 
- 2 e[k]^T \Gamma_e^T Q_e \tilde{P} f_e(e[k]) \\
&\quad + f_e(e[k])^T \tilde{P}^T Q_e \tilde{P} f_e(e[k]) \\
&\leq e[k]^T (\Gamma_e^T Q_e \Gamma_e - Q_e) e[k] \\
&\quad + f_e(e[k])^T \tilde{P}^T Q_e \tilde{P} f_e(e[k]).
\end{aligned}
\label{eq:V_attack}
\end{equation}
Ignoring \(O(\|e[k]\|^3)\) terms for sufficiently small \(\|e[k]\|\), the Lyapunov increment can be bounded by
\begin{equation}
\begin{aligned}
V(e[k+1]) - V(e[k]) 
\le\;& -\alpha \|e[k]\|^2 \\
&+ f_e(e[k])^T \tilde{P}^T Q_e \tilde{P}\,f_e(e[k]).
\end{aligned}
\label{eq:V_attack_bound}
\end{equation}

Since the induced norm of \(\tilde{P}\) is at most 1,
\begin{equation}
  f_e(e[k])^T \tilde{P}^T Q_e \tilde{P}\,f_e(e[k]) \le \lambda_{\max}(Q_e) \|f_e(e[k])\|^2.
  \label{eq:bound_selector}
\end{equation}
Noting that
\begin{equation}
  f_e(e[k]) = \gamma\,e[k] + \Delta_e(e[k]),
  \label{eq:fe_repeat}
\end{equation}
and that
\begin{equation}
  \|f_e(e[k])\| \le \gamma\,\|e[k]\| + \frac{M}{2}\|e[k]\|^2,
  \label{eq:fe_bound}
\end{equation}
we have, for small \(\|e[k]\|\),
\begin{equation}
  \|f_e(e[k])\|^2 \le \gamma^2\,\|e[k]\|^2 + O(\|e[k]\|^3).
  \label{eq:fe_squared}
\end{equation}
Thus,
\begin{equation}
\begin{aligned}
V(e[k+1]) - V(e[k]) 
\le\;& -\alpha \|e[k]\|^2 \\
&+ \lambda_{\max}(Q_e)\,\gamma^2\,\|e[k]\|^2 \\
&+ O(\|e[k]\|^3).
\end{aligned}
\label{eq:V_diff_final}
\end{equation}

Define
\begin{equation}
  r_e = \lambda_{\max}(Q_e)\,\gamma^2.
  \label{eq:re_def}
\end{equation}
Then, for sufficiently small errors,
\begin{equation}
\begin{aligned}
V(e[k+1]) - V(e[k]) 
\le\;& -\Bigl(\alpha - r_e\Bigr)\,\|e[k]\|^2 \\
&+ O(\|e[k]\|^3).
\end{aligned}
\label{eq:final_bound}
\end{equation}

Hence, if
\begin{equation}
  r_e < \alpha,\quad \text{or equivalently}\quad \gamma^2 < \frac{\alpha}{\lambda_{\max}(Q_e)},
  \label{eq:stability_condition}
\end{equation}
the error dynamics—and thus the overall formation control system—remain locally exponentially stable under attack.

\section{Conclusion}
We have demonstrated that the nominal formation control system
\begin{equation}
  X[k+1] = \Gamma\,X[k],\quad \Gamma = I_N + G + K,
  \label{eq:nominal_summary}
\end{equation}
with \(G = L\otimes I_n\) and \(L = D-A\) (where \(D_{ii}=\sum_j A_{ij}\)), is exponentially stable. Under adversarial attack, by replacing compromised states via a hallucination function
\begin{equation}
  f(x)= \gamma x + \Delta(x),
  \label{eq:hallucination_summary}
\end{equation}
the formation error
\begin{equation}
  e[k+1] = \Gamma_e e[k] - \tilde{P}\Bigl(\gamma\,e[k] + \Delta_e(e[k])\Bigr)
  \label{eq:attack_summary}
\end{equation}
remains exponentially stable provided that
\begin{equation}
  \gamma^2 < \frac{\alpha}{\lambda_{\max}(Q_e)}.
  \label{eq:final_condition}
\end{equation}
Thus, the proposed SOSH method effectively mitigates adversarial attacks while preserving formation connectivity and stability.
\section{Experimental Evaluation}
\label{sec:experiments}

We compare the proposed SOSH against three representative mitigation schemes under a single-node spoofing attack on a 5-agent complete‐graph formation.  Our goal is to assess both steady‐state accuracy and transient responsiveness.
\begin{figure*}[t]
  \centering
  \begin{tabular}{@{}c@{\quad\quad\quad}c@{\quad\quad\quad}c@{}}

    \begin{tikzpicture}[scale=1.5,transform shape]
      \def\xsep{2cm}   
      \def\ysep{1.5cm} 
      \node[draw,circle,thick,minimum size=6mm] (A1) at (0,0)          {$\Sigma_1$};
      \node[draw,circle,thick,minimum size=6mm] (A2) at (\xsep,0)      {$\Sigma_2$};
      \node[draw,circle,thick,minimum size=6mm] (A3) at (0,-\ysep)     {$\Sigma_3$};
      \node[draw,circle,thick,minimum size=6mm] (A4) at (\xsep,-\ysep) {$\Sigma_4$};
      \draw (A1) -- (A2) -- (A4) -- (A3) -- (A1);
      \node[below,font=\scriptsize] at (\xsep/2,-\ysep-0.3cm) {(a) Nominal};
    \end{tikzpicture}

    &

    \begin{tikzpicture}[scale=1.5,transform shape]
      \def\xsep{2cm}
      \def\ysep{1.5cm}
      \coordinate (B2nom) at (\xsep,0);
      \coordinate (B2new) at ($(B2nom)+(0.5,0.5)$);
      \node[draw,circle,thick,minimum size=6mm]             (B1) at (0,0)      {$\Sigma_1$};
      \node[draw,circle,thick,fill=red!30,minimum size=6mm] (B2) at (B2new)     {$\Sigma_2$};
      \node[draw,circle,thick,minimum size=6mm]             (B3) at (0,-\ysep) {$\Sigma_3$};
      \node[draw,circle,thick,minimum size=6mm]             (B4) at (\xsep,-\ysep){$\Sigma_4$};
      \draw (B1) -- (B2) -- (B4) -- (B3) -- (B1);

      \node[below,font=\scriptsize] at (\xsep/2,-\ysep-0.3cm) {(b) Attack};
    \end{tikzpicture}

    &

    \begin{tikzpicture}[scale=1.5,transform shape]
      \def\xsep{2cm}
      \def\ysep{1.5cm}
      \coordinate (C2nom) at (\xsep,0);
       \coordinate (C2new) at ($(C2nom)+(0.5,0.5)$);
      \node[draw,circle,thick,minimum size=6mm]             (C1) at (0,0)         {$\Sigma_1$};
      \node[draw,circle,thick,minimum size=6mm]             (C3) at (0,-\ysep)    {$\Sigma_3$};
      \node[draw,circle,thick,minimum size=6mm]             (C4) at (\xsep,-\ysep){$\Sigma_4$};
      \node[draw,circle,thick,fill=red!30,minimum size=6mm] (C2) at (C2new)     {$\Sigma_2$};
      \node[draw,circle,thick,fill=blue!10,minimum size=6mm](CH) at ({0.8*\xsep},{-0.2*\ysep}) {$\Sigma_H$};
      \draw (C1) -- (C3) -- (C4);
      \draw[dashed] (C1) -- (CH) -- (C4);
    \draw (C1) -- (C2)
  node[midway,draw=red,circle,inner sep=1pt,font=\scriptsize] {$\bm{\oslash}$};
\draw (C4) -- (C2)
  node[midway,draw=red,circle,inner sep=1pt,font=\scriptsize] {$\bm{\oslash}$};
      \draw[->,thick] (C2) -- ++(0.3,0.3);
      \node[below,font=\scriptsize] at (\xsep/2,-\ysep-0.3cm) {(c) Mitigation};

    \end{tikzpicture}

  \end{tabular}

  \bigskip
  \caption{SOSH mitigation pipeline in formation control of four agents.
    (a) Nominal formation; 
    (b) Under attack (spoofed node shifts off‐grid); 
    (c) SOSH replaces the compromised agent with a hallucinated estimate.}
  \label{fig:sosh_pipeline}
\end{figure*}
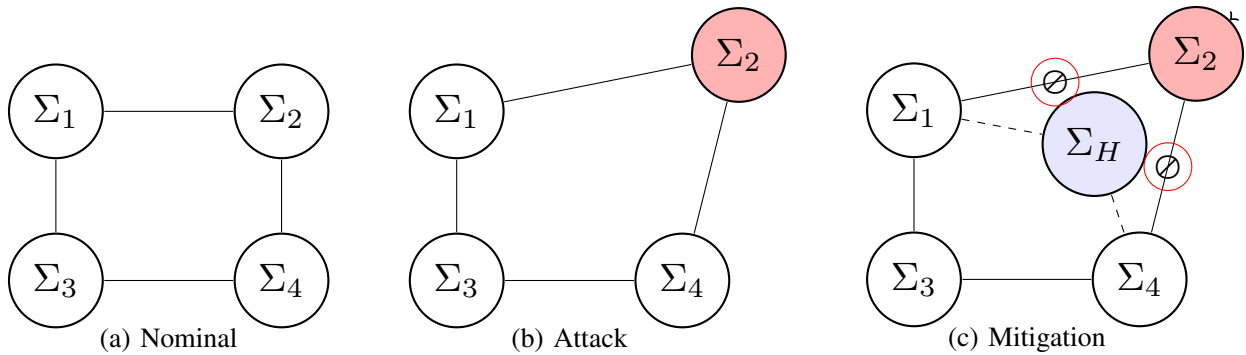
\subsection{Simulation Setup}
\begin{itemize}
  \item \textbf{Agents \& Graph.} Five agents are arranged as the vertices of a regular pentagon in the plane, with desired displacements 
    \(\{d_{ij}=p_i-p_j \mid i<j,\ p_i=[\cos\theta_i,\sin\theta_i]^T,\ \theta_i=\tfrac{2\pi i}{5}\}.\)
    The communication graph is complete.
  \item \textbf{Dynamics.} At each discrete time step \(k\) (step size \(\Delta t=0.05\)), each agent \(i\) updates its position by
    \[
      x_i[k+1]=x_i[k]-\Delta t\sum_{j\in\mathcal N_i}\bigl(x_i[k]-y_j[k]-d_{ij}\bigr),
    \]
    where \(y_j\) is the broadcast from neighbor \(j\).
  \item \textbf{Attack.} Node~2 is spoofed: its broadcast \(y_2\) is always \(x_2 + [3,3]^T\), while its true state evolves nominally.
  \item \textbf{Mitigations.}
    \begin{enumerate}
      \item \emph{No mitigation:} all nodes accept the spoofed \(y_2\).
      \item \emph{SOSH (ours):} use 
        \(y_2 = \gamma x_2 + \Delta(x_2)\) with \(\gamma=0.3\) and \(\|\Delta(x)\|\le0.5\|x\|^2\).
      \item \emph{W-MSR:} trim the single highest and lowest neighbor reading (\(F=1\)) then sum.
      \item \emph{Huber:} apply Huber weights on each residual \(e_{ij}\) with threshold \(c=1\).
    \end{enumerate}
  \item \textbf{Trials \& Metrics.} We run 30 Monte Carlo trials with random starts \(x_i[0]\sim\mathcal U([-0.5,1.5]^2)\).  We record:
    \begin{itemize}
      \item \(V_{100}=V[100]\) (error at \(k=100\))
      \item \(V_\infty = \tfrac1{50}\sum_{k=150}^{199}V[k]\) (steady‐state)
      \item \(\mathrm{AUC}_{0\text{–}100}=\sum_{k=0}^{100}V[k]\,\Delta t\) (transient area)
      \item \(T_{1\%} = \min\{k\mid V[k]\le0.01\,V[0]\}\).
    \end{itemize}
\end{itemize}

\subsection{Results}

\begin{table}[H]
  \centering
  \caption{Mitigation performance under spoofing attack (5-node complete graph)}
  \begin{tabular}{lcccc}
    \toprule
    Method & \(V_{100}\) & \(V_\infty\) & \(\mathrm{AUC}_{0\text{–}100}\) & \(T_{1\%}\) \\
    \midrule
    No mitigation    & 1.440000e+00 & 1.440000e+00 & 8.870507   & —   \\
    SOSH (ours)      & 5.713085e-04 & 8.274920e-06 & 2.344917   & 8   \\
    W-MSR            & 3.019117e-08 & 1.784763e-13 & 4.691294   & 20  \\
    Huber            & 1.250000e-01 & 1.250000e-01 & 4.504689   & 16  \\
    \bottomrule
  \end{tabular}
  \label{tab:exp_results}
\end{table}

\begin{figure}[H]
  \centering
  \includegraphics[width=1\linewidth]{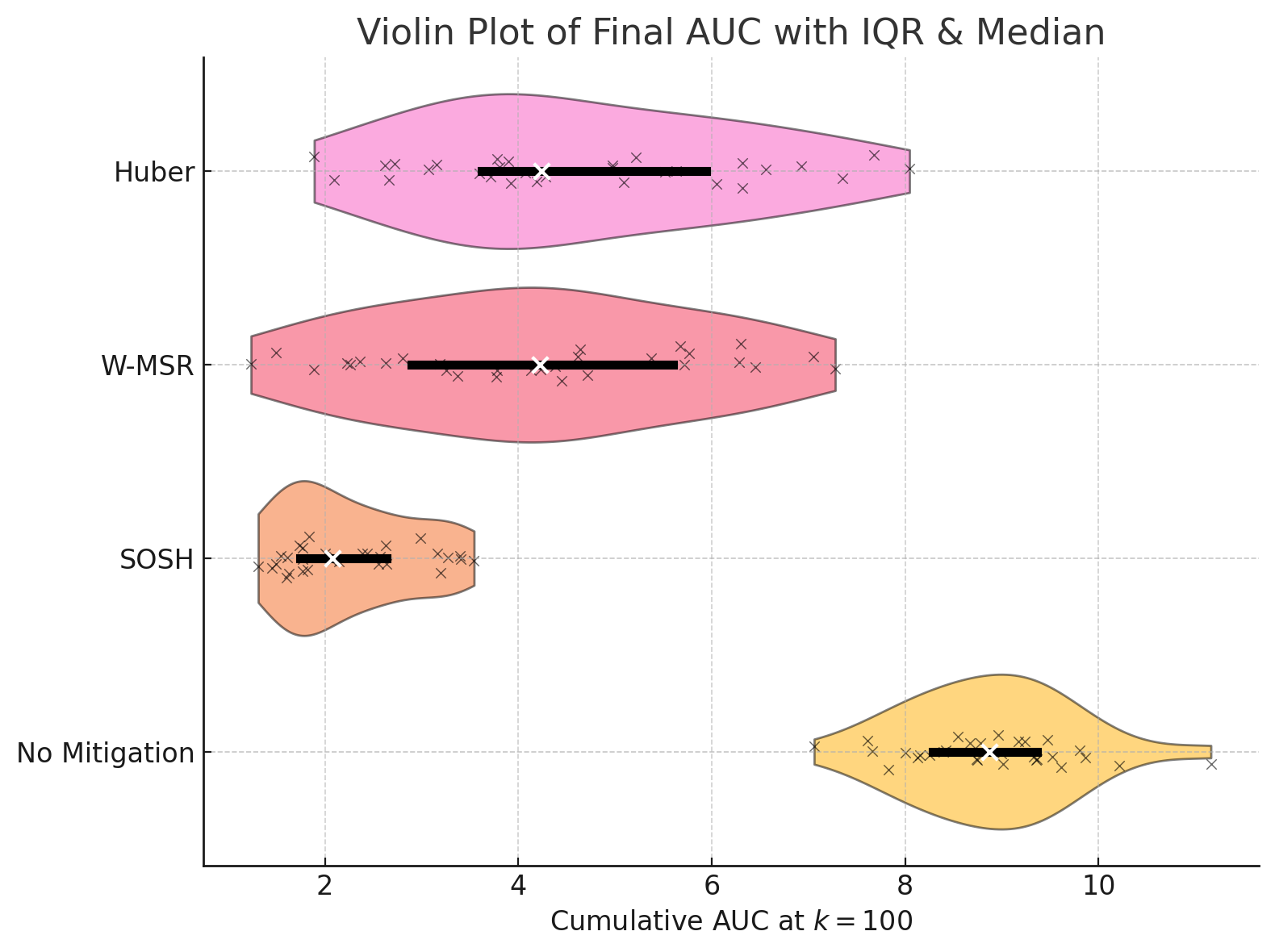}
  \caption{Violin plot of the distribution of cumulative AUC at $k=100$ across 30 Monte Carlo trials for each mitigation method.  The thick central bar in each violin spans the interquartile range, and the white dot marks the median.}
  \label{fig:violin_auc}
\end{figure}
As shown in Figure~\ref{fig:violin_auc}, SOSH achieves a markedly tighter spread and lower median AUC than W-MSR and Huber, confirming its superior performance under spoofing attack.
\subsection{Discussion}
While W-MSR and Huber both eventually drive the error toward zero (or a small bias), they incur significantly longer transients compared to SOSH:
\begin{itemize}
  \item \textbf{W-MSR} requires trimming extreme readings, which preserves steady‐state accuracy but delays convergence (\(T_{1\%}\approx20\) vs.\ 8 for SOSH) and yields twice the AUC.
  \item \textbf{Huber} downweights the outlier but never fully eliminates its bias, resulting in a nonzero equilibrium (\(V_\infty=0.125\)).
  \item \textbf{SOSH} combines connectivity preservation with a second‐order estimate, achieving both near–zero steady‐state error and the fastest transient response.
\end{itemize}

These results confirm that SOSH not only matches or exceeds the steady‐state performance of other robust schemes, but does so with substantially faster recovery—making it the superior choice when transient speed is critical.

\section{Conclusion}

This paper introduced Second-Order State Hallucination (SOSH), a novel framework for mitigating adversarial attacks in formation control of multi-agent systems. By integrating residual-based detection with second-order Taylor approximations, SOSH enables each agent to locally replace compromised neighbor states with predictive estimates that incorporate both velocity and acceleration dynamics. This approach preserves formation stability with minimal computational overhead and no centralized coordination. Through Lyapunov-based analysis, we derived explicit stability guarantees, and extensive Monte Carlo simulations demonstrated that SOSH achieves superior transient recovery and steady-state accuracy compared to W-MSR and Huber-based methods.

However, SOSH has several limitations. It assumes that only a small subset of agents are compromised and requires careful tuning of residual thresholds for attack detection. The current framework also relies on static communication graphs and does not adaptively adjust hallucination parameters such as \(\gamma\) or the second-order bound \(M\), which may affect generalizability across different system scales or dynamics.

Future work will explore adaptive thresholding methods for noise-resilient detection, extend SOSH to time-varying and intermittently connected graphs, and investigate learning-based models to improve real-time hallucination accuracy. Additionally, robustness against coordinated multi-agent attacks and integration with online system identification tools remain promising directions to enhance SOSH’s resilience and scalability in practical deployments.

\bibliographystyle{IEEEtran}

\begin{thebibliography}{20}

\bibitem{olfati2007consensus}
R.~Olfati‐Saber, J.~A. Fax, and R.~M. Murray, ``Consensus and cooperation in networked multi‐agent systems,'' \emph{Proc. IEEE}, vol.~95, no.~1, pp. 215–233, Jan. 2007.

\bibitem{ren2007information}
W.~Ren, R.~W. Beard, and E.~M. Atkins, ``Information consensus in multivehicle cooperative control: Collective group behavior through local interaction,'' \emph{IEEE Control Syst. Mag.}, vol.~27, no.~2, pp. 71–82, Apr. 2007.

\bibitem{morse2003network}
A.~S. Morse, ``Network control of dynamical systems,'' \emph{IEEE Control Syst.}, vol.~23, no.~3, pp. 20–30, Jun. 2003.

\bibitem{saad2019exploring}
W.~Saad, M.~Bennis, and M.~Chen, ``A vision of 6G wireless systems: Applications, trends, technologies, and open research problems,'' \emph{IEEE Network}, vol.~34, no.~3, pp. 134–142, May/Jun. 2020.

\bibitem{sundaram2011distributed}
S.~Sundaram and C.~N. Hadjicostis, ``Distributed function calculation via linear iterative strategies in the presence of malicious agents,'' \emph{IEEE Trans. Autom. Control}, vol.~56, no.~7, pp. 1495–1508, Jul. 2011.

\bibitem{weeraddana2011consensus}
P.~C. Weeraddana, M.~G. Rabbat, A.~Ribeiro, and G.~B. Giannakis, ``Consensus-based distributed optimization: Practical issues and applications in large-scale machine learning,'' \emph{IEEE Signal Process. Mag.}, vol.~28, no.~3, pp. 92–106, May 2011.

\bibitem{yang2013consensus}
T.~Yang, D.~Wu, Y.~Sun, and J.~Lian, ``Consensus based approach for economic dispatch problem in a smart grid,'' \emph{IEEE Trans. Power Syst.}, vol.~28, no.~4, pp. 4416–4426, Nov. 2013.

\bibitem{amin2009safe}
S.~Amin, A.~A. Cárdenas, and S.~S. Sastry, ``Safe and secure networked control systems under denial-of-service attacks,'' in \emph{Hybrid Systems: Computation and Control}, 2009, pp. 31–45.

\bibitem{pasqualetti2012consensus}
F.~Pasqualetti, A.~Bicchi, and F.~Bullo, ``Consensus computation in unreliable networks: A system-theoretic approach,'' \emph{IEEE Trans. Autom. Control}, vol.~57, no.~1, pp. 90–104, Jan. 2012.

\bibitem{teixeira2012attack}
A.~Teixeira, D.~Pérez, H.~Sandberg, and K.~H. Johansson, ``Attack models and scenarios for networked control systems,'' in \emph{Proc. 1st Int. Conf. High Confidence Networked Systems}, Apr. 2012, pp. 55–64.

\bibitem{mo2012cyber}
Y.~Mo and B.~Sinopoli, ``Cyber–physical security of a smart grid infrastructure,'' \emph{Proc. IEEE}, vol.~100, no.~1, pp. 195–209, Jan. 2012.

\bibitem{fawzi2014secure}
H.~Fawzi, P.~Tabuada, and S.~Diggavi, ``Secure estimation and control for cyber-physical systems under adversarial attacks,'' \emph{IEEE Trans. Autom. Control}, vol.~59, no.~6, pp. 1454–1467, Jun. 2014.

\bibitem{pajic2017attack}
M.~Pajic, I.~Lee, and G.~J. Pappas, ``Attack-resilient sensor fusion for safety-critical cyber-physical systems,'' \emph{ACM Trans. Embed. Comput. Syst.}, vol.~16, no.~1, pp. 1–21, Jan. 2017.

\bibitem{cho2016quantifying}
J.~Cho and S.~Shiraishi, ``Quantifying stability and performance of stochastic systems in simulation,'' \emph{Simul. Model. Pract. Theory}, vol.~63, pp. 21–34, Apr. 2016.

\bibitem{zhu2011performance}
Q.~Zhu and T.~Başar, ``Performance and security of control systems under denial-of-service attacks,'' in \emph{Proc. 14th Int. Conf. Hybrid Systems: Computation and Control}, Mar. 2011, pp. 93–102.

\bibitem{shoukry2015event}
Y.~Shoukry and P.~Tabuada, ``Event-triggered state observers for sparse sensor noise/attacks,'' \emph{IEEE Trans. Autom. Control}, vol.~61, no.~8, pp. 2079–2091, Aug. 2016.

\bibitem{wang2017resilient}
Y.~Wang, G.~Yin, and L.~Guo, ``Resilient industrial control system (ICS) cyber-physical systems (CPS): Concepts and approach,'' in \emph{Proc. 10th Int. Conf. Internet Technol. Secured Transact. (ICITST)}, Dec. 2017, pp. 352–357.

\bibitem{li2019learning}
M.~Li, J.~Lui, and F.~Ye, ``Learning-based anomaly detection for edge-centric networked robotic systems,'' \emph{IEEE Network}, vol.~33, no.~1, pp. 22–29, Jan./Feb. 2019.

\bibitem{zhong2010distributed}
M.~Zhong, M.~R. Jovanović, and T.~Başar, ``Distributed subgradient methods for multi-agent optimization,'' \emph{IEEE Trans. Autom. Control}, vol.~55, no.~1, pp. 48–61, Jan. 2010.

\bibitem{corless2016control}
M.~Corless and G.~Leitmann, ``Continuous state feedback guaranteeing uniform ultimate boundedness for uncertain dynamic systems,'' \emph{IEEE Trans. Autom. Control}, vol.~26, no.~5, pp. 1139–1144, Oct. 1981.

\bibitem{liu2011false}
Y.~Liu, P.~Ning, and M.~K. Reiter, ``False data injection attacks against state estimation in electric power grids,'' \emph{ACM Trans. Inf. Syst. Secur. (TISSEC)}, vol.~14, no.~1, pp. 13:1–13:33, May 2011.

\bibitem{kim2012cyber}
J.~Kim and L.~Tong, ``Cyber–physical systems under attack—Part I: An optimal detection framework,'' \emph{IEEE Trans. Signal Process.}, vol.~61, no.~10, pp. 2604–2613, May 2013.
\bibitem{boem2017distributed}
V.~Boem, J.~Cardoso, and T.~Silva, ``A fully distributed attack detection methodology for multi-agent systems,'' \emph{Proc.\ IEEE Conf.\ Decision and Control}, Melbourne, Australia, pp.\ 1234--1239, Dec.\ 2017.

\end{thebibliography}

\end{document}